\documentstyle[graphics,psfig]{mn}

\newif\ifAMStwofonts
\ifoldfss
  \ifCUPmtlplainloaded \else
    \NewTextAlphabet{textbfit} {cmbxti10} {}
    \NewTextAlphabet{textbfss} {cmssbx10} {}
    \NewMathAlphabet{mathbfit} {cmbxti10} {} % for math mode
    \NewMathAlphabet{mathbfss} {cmssbx10} {} %  "   "    "
  \fi
  \ifAMStwofonts
    \ifCUPmtlplainloaded \else
      \NewSymbolFont{upmath} {eurm10}
      \NewSymbolFont{AMSa} {msam10}
      \NewMathSymbol{\upi}     {0}{upmath}{19}
      \NewMathSymbol{\umu}     {0}{upmath}{16}
      \NewMathSymbol{\upartial}{0}{upmath}{40}
      \NewMathSymbol{\leqslant}{3}{AMSa}{36}
      \NewMathSymbol{\geqslant}{3}{AMSa}{3E}

    \fi
  \fi
\fi % End of OFSS

\ifnfssone
  \newmathalphabet{\mathit}
  \addtoversion{normal}{\mathit}{cmr}{m}{it}
  \addtoversion{bold}{\mathit}{cmr}{bx}{it}
  \newmathalphabet{\mathbfit} % math mode version of \textbfit{..}
  \addtoversion{normal}{\mathbfit}{cmr}{bx}{it}
  \addtoversion{bold}{\mathbfit}{cmr}{bx}{it}
  \newmathalphabet{\mathbfss} % math mode version of \textbfss{..}
  \addtoversion{normal}{\mathbfss}{cmss}{bx}{n}
  \addtoversion{bold}{\mathbfss}{cmss}{bx}{n}
  \ifAMStwofonts
    \ifCUPmtlplainloaded \else
      \UseAMStwoboldmath
      \makeatletter
      \new@mathgroup\upmath@group
      \define@mathgroup\mv@normal\upmath@group{eur}{m}{n}
      \define@mathgroup\mv@bold\upmath@group{eur}{b}{n}
      \edef\UPM{\hexnumber\upmath@group}
      \new@mathgroup\amsa@group
      \define@mathgroup\mv@normal\amsa@group{msa}{m}{n}
      \define@mathgroup\mv@bold\amsa@group{msa}{m}{n}
      \edef\AMSa{\hexnumber\amsa@group}
      \makeatother
      \mathchardef\upi="0\UPM19
      \mathchardef\umu="0\UPM16
      \mathchardef\upartial="0\UPM40
      \mathchardef\leqslant="3\AMSa36
      \mathchardef\geqslant="3\AMSa3E
    \fi
  \fi
\fi % End of NFSS release 1

\ifnfsstwo
  \DeclareMathAlphabet{\mathbfit}{OT1}{cmr}{bx}{it}
  \SetMathAlphabet\mathbfit{bold}{OT1}{cmr}{bx}{it}
  \DeclareMathAlphabet{\mathbfss}{OT1}{cmss}{bx}{n}
  \SetMathAlphabet\mathbfss{bold}{OT1}{cmss}{bx}{n}
  \ifAMStwofonts
    \ifCUPmtlplainloaded \else
      \DeclareSymbolFont{UPM}{U}{eur}{m}{n}
      \SetSymbolFont{UPM}{bold}{U}{eur}{b}{n}
      \DeclareSymbolFont{AMSa}{U}{msa}{m}{n}
      \DeclareMathSymbol{\upi}{0}{UPM}{"19}
      \DeclareMathSymbol{\umu}{0}{UPM}{"16}
      \DeclareMathSymbol{\upartial}{0}{UPM}{"40}
      \DeclareMathSymbol{\leqslant}{3}{AMSa}{"36}
      \DeclareMathSymbol{\geqslant}{3}{AMSa}{"3E}
    \fi
  \fi
\fi % End of NFSS release 2

\ifCUPmtlplainloaded \else
  \ifAMStwofonts \else % If no AMS fonts
    \def\upi{\pi}
    \def\umu{\mu}
    \def\upartial{\partial}
  \fi
\fi

\def\pc{\,{\rm pc}}

\def\Mpc{\,{\rm Mpc}}
\def\msun{\,{\rm M}_\odot}
\def\kms{\,{\rm km}\,{\rm s}^{-1}}
\def\etal{et al.\ }
\def\PA{\hbox{PA}}
\def\eg{{\it e.g.\ }}
\def\ie{{\it i.e.\ }}

\title[Kinematics from wide slits]
{Kinematics from spectroscopy with a wide slit: \\ 
detecting black holes in galaxy centres}

\author[Witold Maciejewski and James Binney]
{Witold Maciejewski and James Binney\\
Theoretical Physics, 1 Keble Road, University of Oxford, Oxford, OX1 3NP}

\begin{document}

\maketitle

\begin{abstract}
We consider long-slit emission-line spectra of galactic nuclei when the slit
is wider than the instrumental PSF, and the target has large velocity
gradients. The finite width of the slit generates complex distributions of
brightness at a given spatial point in the measured spectrum, which can be
misinterpreted as coming from additional physically distinct nuclear
components.  We illustrate this phenomenon for the case of a thin disc in
circular motion around a nuclear black hole (BH). We develop a new method
for estimating the mass of the BH that exploits a feature in the spectrum at
the outer edge of the BH's sphere of influence, and therefore gives higher
sensitivity to BH detection than traditional methods. Moreover, with this
method we can determine the black hole mass and the inclination of the
surrounding disc separately, whereas the traditional approach to black-hole
estimation requires two long-slit spectra to be taken. We show that with a
given spectrograph, the detectability of a BH depends on the sense of
rotation of the nuclear disc. We apply our method to estimate the BH mass in
M84 from a publicly available spectrum, and recover a value 4 times lower than
that published previously from the same data.
 \end{abstract}


\section{Introduction}

Long-slit spectroscopy has been for decades a fundamental tool of astronomy,
and although integral-field spectrographs are now beginning to be important 
for ground-based observations of galactic dynamics (e.g., SAURON, Copin \etal 
2000), the Hubble Space Telescope (HST) has never been more reliant on 
long-slit spectroscopy.

Observations with long-slit spectrographs often use a slit 
that is wider than the FWHM of the instrumental point-spread-function (PSF). 
This practice is conventionally considered to enhance the signal-to-noise 
ratio (S/N) of the data at the price of what may be an insignificant loss in
velocity resolution. Here we show that when the target has steep velocity
gradients, the use of a wide slit can have more subtle effects, because
the position and velocity information becomes entangled along the dispersion 
direction. If these
effects are not recognized in the data, a misleading impression of the
structure of the target can be inferred.  When the effects of
a wide slit are recognized, however, they can increase the diagnostic power
of the spectrum over that of a narrow-slit spectrum of equal S/N.

Since our study has been motivated by a programme of spectroscopy of
galactic nuclei using the STIS instrument on HST (Marconi \etal 2000;
Axon \etal 2000), we focus on the specific
case of STIS spectra of gas swirling around a black hole in a galactic
nucleus. However, the principles we elucidate have a wider applicability:
whenever an object with steep velocity gradients is studied with a slit that
is wider than the FWHM of the PSF. Large velocity gradients occur in shocks
and contact discontinuities, as well as in accretion discs around black
holes.

In Section 2 we analyze the appearance of galactic-disc spectra that are 
taken with a slit that
is much wider than the PSF and has been laid along the disc's line of nodes.
In Section 3 we use this analysis to develop a new estimator for the mass of
the nuclear BH. In Section 4 we generalize the analysis of Sections 2 and 3
to a slit that is inclined to the line of nodes. We find
that, paradoxically, the detectability of a BH with a given spectrograph
depends on the sense of rotation of its accretion disc. We also investigate
the case of a slit that does not pass through the nucleus. In Section 5 we 
consider the case in which the FWHM of the PSF is comparable to the slit 
width. In Section 6 we
reanalyse STIS observations of M84 taken by Bower et al (1998), and show
that the results of Section 4 enable us to understand these data with a much
simpler model of the galaxy than that developed by Bower et al., and yield a
mass for the black hole in M84 that is smaller than that of Bower et al.\ by
a factor $\sim4$.

\section{A slit wider than the PSF}

For clarity we consider long-slit spectra of a disc in circular motion in a
specific but realistic axisymmetric galactic potential. We assume that the
latter is generated by a nuclear BH and an extended mass distribution, in
which the density varies radially as $R^{-1.8}$ -- this behaviour is
consistent with what is inferred for the centre of the Milky Way (Binney et
al., 1991; Genzel et al., 1997) and is characteristic of the luminosity
profiles of a large class of elliptical galaxies (Gebhardt et al., 1996).
With the density scaling as $R^{-1.8}$, the circular speed generated by the
extended mass distribution scales as $R^{0.1}$, and falls towards the
centre. Consequently, there is some radius $R_{\rm BH}$ at which the
circular speed of the embedded BH, which scales as $R^{-0.5}$,
begins to dominate.
The conventional approach to the determination of the mass of the BH
involves fitting a Keplerian circular-speed curve to the observed rotation
curve well inside $R_{\rm BH}$. We shall see that with a wide slit 
mass inside $R_{\rm BH}$ can be derived from data taken at $R\simeq
R_{\rm BH}$. The ability to work near $R_{\rm BH}$ rather than well inside
it is important  because $R_{BH}$ is typically very small --- usually smaller
than atmospheric seeing.

\begin{figure}
\centerline{\psfig{file=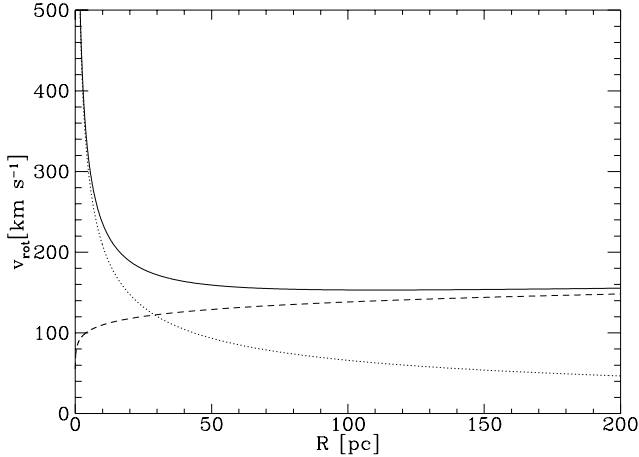,width=\hsize,angle=-90}}
 \caption{Circular-speed curves for the inner $200\pc$ of our model disc:
the {\it dashed line} is the contribution of the stars, which is assumed to
rise as $R^{0.1}$; the {\it dotted line} is the contribution from a
$10^8\msun$ nuclear BH; the {\it solid line} is the net circular speed.
\label{rotcurve}}
\end{figure}

Fig.~\ref{rotcurve} shows the circular speeds out to $200\pc$ in our model:
the dotted curve shows the circular speed generated by the $10^8\msun$ BH, 
the dashed line shows that generated by the stars, and the full curve shows 
the net circular speed. 

\subsection{A slit parallel to the line of nodes}

\begin{figure}
\centerline{\psfig{file=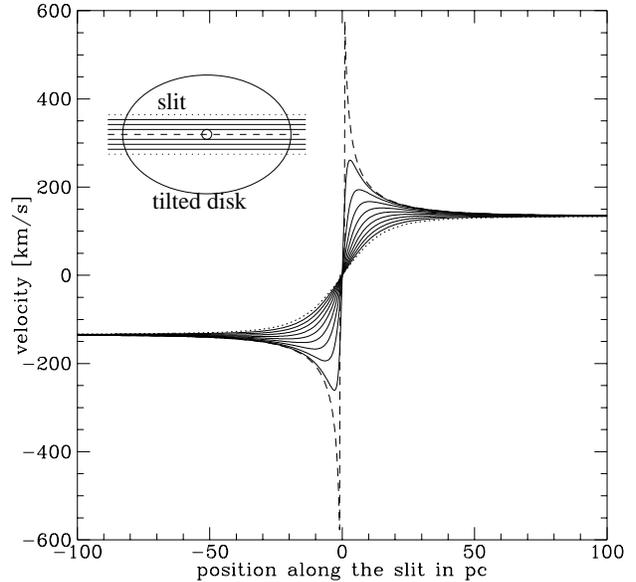,width=\hsize}}
 \caption{Position-velocity diagram for the disc described by
Fig.~\ref{rotcurve} when it is viewed at inclination angle $i=60\degr$.
Velocities are sampled along 21 cuts parallel to the line of nodes. They
are symmetric with respect to the nucleus, hence only 11 lines are
displayed. The cuts are represented in the inset at upper left, with the cut
through the nucleus dashed, and the outermost cut, which passes
$20\cos^{-1}i\pc$ from the nucleus, dotted.
\label{larger}}
\end{figure}

Fig.~\ref{larger} is a composite position-velocity plot for 21 parallel
cuts through our model galaxy. The disc has been inclined at $60\degr$ to the
line of sight, and the cuts run parallel to the line of nodes. The inset at 
top left shows the locations of some representative
cuts: the outermost cuts are dotted and the one through the nucleus is dashed. 
Velocities along the cuts differ markedly near the centre since only
the cut through the nucleus fully samples the central divergence of the BH's
circular-speed  curve. Velocities along many outer cuts show no rise, 
but smoothly decline to zero at the slit centre.

\subsection{Velocity offsets}
When our disc is observed in some emission line through a long-slit
spectrograph, the pattern of intensity resembles Fig.~\ref{larger} but with
some significant differences, which we now deduce by considering how the
spectrograph works.

\begin{figure}
\centerline{\psfig{file=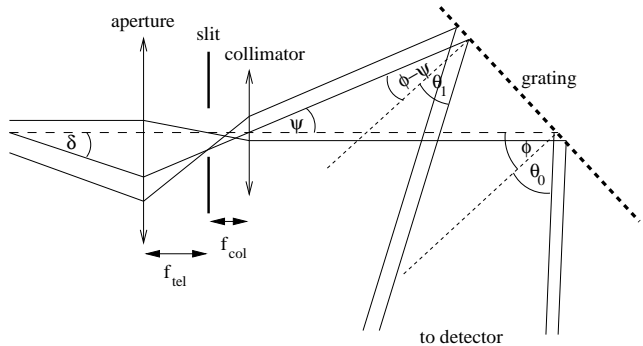,width=\hsize}}
 \caption{Paths through an ideal spectrograph of rays associated with two
points on the image of an astronomical object.  One ray runs along the
optical axis (long-dashed line), and contributes to a maximum intensity in
the direction $\theta_0$ relative to the grating. A ray through an image
point that lies distance $\delta$ from the axis contributes
to an intensity maximum at angle $\theta_1$ relative to the grating.
Short-dashed lines mark normals to the grating.\label{spectrograph}}
\end{figure}

Fig.~\ref{spectrograph} shows the relevant geometry. The telescope creates
an image of the disc in the slit, and we show two rays of wavelength
$\lambda$ through each of two points on this image: one at the centre of the 
slit, which lies on the optical axis, and one off-centre. After passage through 
the collimator, rays through the centre of the slit run parallel to the optical
axis until they hit the grating, whose normal makes angle $\phi$ with the
axis. The interference pattern produced by the grating peaks at angle
$\theta_0$ to the grating's normal, where $\theta_0$ satisfies the
diffraction equation

\begin{equation}\label{givest0}
\sin\theta_0 - \sin \phi =  m \lambda / \Delta,
\end{equation}
with $m$ being the spectral order, and $\Delta$ the row separation.

The rays going through a point on the image that is offset by
$\delta$ from the centre of the slit are inclined at angle $\psi$ to the 
optical axis after passing the collimator, where

\begin{equation}
\psi = \delta \frac{f_{\rm telescope}}{f_{\rm collimator}}.
\end{equation}
 Hence, their angle of incidence on the grating is $\phi-\psi$, and the
diffraction pattern to which they contribute peaks at the angle $\theta_1$
that satisfies

\begin{equation}\label{givest1}
\sin \theta_1 - \sin (\phi - \psi) =  m \lambda / \Delta.
\end{equation}
The angle $\theta$ of the peak in the diffraction pattern always decreases
with the displacement $\delta$ of the source from the slit centre.

The spectrograph's camera gathers all light that enters in a given
direction onto a single point on the detector. Since the on- and off-centre
rays enter at different angles, they are gathered to
different points on the detector despite their common wavelength. In so far as
we associate a velocity with each position along the dispersion direction,
we have to
recognize that the velocity scale for a point distance $\delta$ from the
centre of the slit is offset from the scale for the centre of the slit by an
amount proportional to $\delta$. This offset depends on wavelength $\lambda$,
and is listed for individual spectrographs as a `plate scale' (see Table 5
in Woodgate \etal 1998 for STIS). This feature of a spectrograph is commonly
known as a slit effect (see \eg Bacon \etal 1995).

\begin{figure}
\centerline{\psfig{file=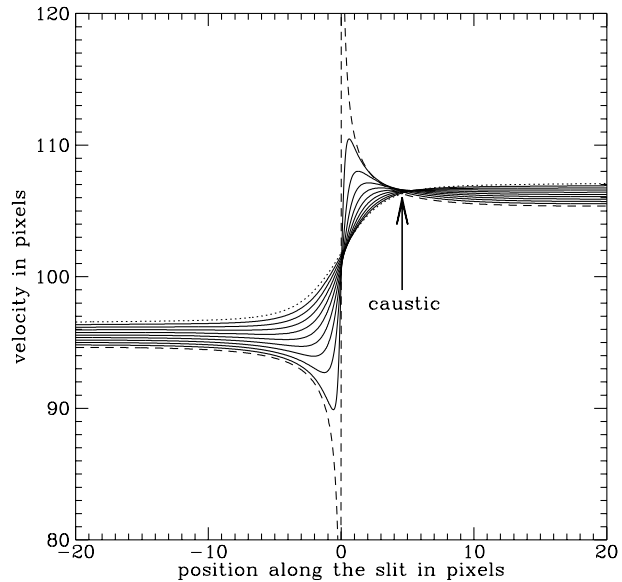,width=\hsize}}
 \caption{The distribution of light on the spectrograph's detector for a
position-velocity diagram like that of Fig.~\ref{larger}. For clarity only
the top half of the slit is sampled -- velocity offsets lift the degeneracy
between the velocities coming from above and below the nucleus in
Fig.~\ref{larger}.  A caustic to the right of the nucleus is marked with an
arrow. The pixel scale is for a slit $0.2\,$arcsec wide and the G750M
grating on STIS with the model disc placed at a distance of
$20\Mpc$.\label{shifts}}
\end{figure}

For the spectral order $m=1$, the angle $\theta$ at which constructive
interference peaks increases with wavelength $\lambda$, or decreases with
emitter's velocity. As noted above, it also decreases as one moves up the
slit in Fig.~\ref{larger}. Out of convenience, we convert decrements in
$\theta$ to increments in the dispersion coordinate on the detector, so that
higher velocity of the emitter reveals itself as larger value of the
dispersion coordinate. With this conversion, also the velocity offset is
positive for points above the centre of the slit, and negative below.
Fig.~\ref{shifts} shows the resulting pattern of light on the detector for
the top half of the slit: each line from Fig.~\ref{larger} is shifted by a
constant amount, different for each line. As in Fig.~\ref{larger}, the
dashed curve comes from the cut through the nucleus, and the dotted curve
marks the edge of the slit, which is $2\delta$ wide. On the left side of
Fig.~\ref{shifts} the dotted curve is shifted up away from the dashed one,
so the light is diffused over the detector in an unhelpful way. On the right
side of Fig.~\ref{shifts}, by contrast, the shift moves the dotted curve
past the dashed curve at large distances along the slit, while at small
distances, the shift leaves the dotted curve below the dashed curve. On this
side, the instrumental velocity offset competes with the real rise in
velocity due to the nuclear BH. Close to the nucleus, the BH-related
velocity rise dominates, hence the dashed line that goes through the nucleus
stays on top. Further from the nucleus, the real velocity gradient
across the slit is small, and the instrumental velocity offset, largest for
the edge of the slit, takes over: hence there the dotted line is on top.
Somewhere in between, there is a point along the slit at which the BH-related
velocity rise is balanced by the velocity offset, and the dashed and dotted
lines intersect.  Since the curves generated by intermediate cuts are
similarly shifted, the dashed curve crosses all the other curves, and
happens to do so in a narrow range of distances along the slit. 
All light entering that part of the slit gets focused virtually
at one position in the dispersion direction. We refer to
this region of inversion of the order of the curves as the `caustic'.

\begin{figure}
\vbox{\qquad\psfig{file=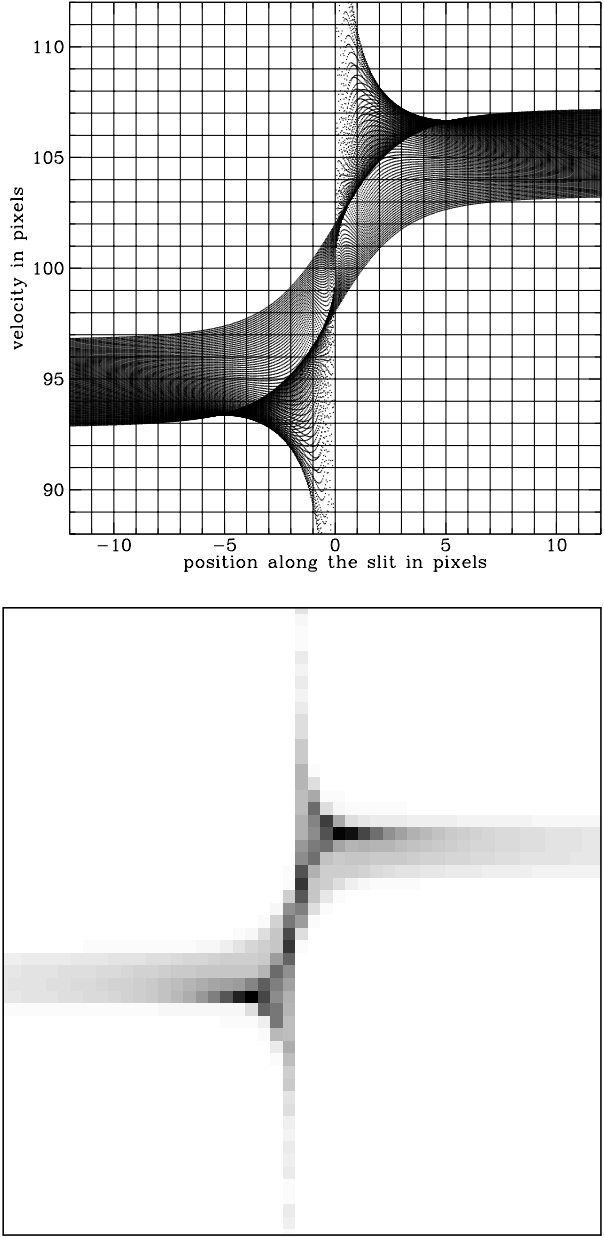,height=\hsize}\qquad
\psfig{file=fig5b.ps,height=\hsize}}
 \caption{The light distribution on the detector of an ideal spectrograph
with a wide slit as in Fig.~\ref{shifts}, but including light from both above
and below the nucleus.  Top panel: light from the disc is sampled at points
uniformly distributed within the slit, and for each point a dot is generated
at the location of maximum intensity in the diffraction pattern.  Bottom
panel: the light distribution when the emission in the nuclear disc scales
with radius as $\sim R^{-1}$ and light has been integrated within the
detector's pixels.\label{all}}
\end{figure}

In the lower half of the slit, the velocity offset is negative, so the light
distribution on the detector from this half of the slit is the same as that
shown in Fig.~\ref{shifts} rotated by $180\degr$. Fig.~\ref{all} shows the
light distribution from the two halves of the slit combined. The top
panel is similar to Fig.~\ref{shifts} except that a grid of pixels has been
overlaid -- the grid is appropriate for the G750M grating on STIS with the
model at a distance $\sim20\Mpc$, so that $1\,$arcsec corresponds to
$100\pc$. The bottom panel shows the light distribution in those pixels
under the assumption that the surface brightness within the disc scales as
$R^{-1}$.

Inside the two caustics, the light distribution shows two maxima in the
dispersion direction: the one furthest from the systemic velocity is created
by light coming from the slit centre, and contains information about the BH.
The other one, made by light from slit edges, goes through zero velocity at
the nucleus. Thus we predict full two-dimensional position-velocity diagrams 
for rotating discs to be rich in structure; the information contained in this
structure is lost if one merely fits a Gaussian emission line profile at each 
radius (which is the traditional approach).

\section{A new BH mass estimator}

Observationally, the caustic described in the last section will be
conspicuous because light from right across the slit is concentrated in it.
We experimented with various stellar mass distributions other than the one
resulting in velocity $\sim R^{0.1}$ presented here, and the position of the
caustic along the slit does not depend sensitively on the stellar density 
profile. Any
reasonable mass distribution produces caustics at locations similar to the
case of no stellar mass at all.  Hence, we can infer the mass of the BH from
the location of the caustics.

The velocity field  in the disc due to the BH is 
\begin{equation}
v = A R^{-1/2},
\end{equation}
where $A$ is related to the BH mass $M_{\bullet}$ by $A=\sqrt{GM_{\bullet}}$. 
Let $(\alpha,\beta)$ be Cartesian coordinates on the sky, with $\beta$ 
constant along the line of nodes, and let $d$ be the distance to the system. 
Then the line-of-sight velocity along a cut offset from the nucleus by 
distance $\beta$ and parallel to the line of nodes is
 \begin{equation}\label{vzero}
v_{\alpha \beta} = \frac{A \ \alpha \ \sin i } 
{d^{1/2}\left(\alpha^2 + ( \frac{\beta}{\cos i})^2 \right) ^{3/4}}.
\end{equation}
Taking the ratio of velocity $v_{\alpha 0}$ at the slit centre ($\beta=0$)
to  that,
$v_{\alpha \delta}$, at the slit top edge ($\beta=\delta$) for a given 
position $\alpha$ along the slit returns the value of inclination angle $i$
\begin{equation}\label{givesi}
\cos i= \frac{\delta}{\alpha \sqrt{(v_{\alpha 0}/v_{\alpha \delta})^{4/3} - 1} } .
\end{equation}
It is easiest to estimate the ratio $v_{\alpha 0}/v_{\alpha \delta}$ at the 
caustic: through a spectrograph of dispersion $D$ (in $\hbox{pixel}/\!\kms$) 
and slit half-width $\delta$, the light from the slit centre is detected at a
position $D v_{\alpha 0}$ in the dispersion direction, and the light from the 
slit top edge is detected at $D v_{\alpha \delta} + B\delta$, where $B$ is 
the detector plate scale in the dispersion direction (in pixel/arcsec). Thus 
although always  $v_{\alpha \delta} < v_{\alpha 0}$, the instrumental velocity
offset $B\delta$, positive for the top half of the slit, moves the light
from the slit edge  up towards the light from the slit centre that is emitted 
at higher intrinsic velocity. If the caustic occurs distance $\alpha$ along 
the slit from the nucleus, one has
 \begin{equation}\label{givesy}
y\equiv D v_{\alpha 0} = D v_{\alpha \delta} + B\delta,
\end{equation}
where $y$ is the position of the caustic in the dispersion direction.
Therefore $v_{\alpha \delta}/v_{\alpha 0} = 1 - \frac{B\delta}{y}$ and
(\ref{givesi}) expressed in observable quantities takes the form
 \begin{equation}\label{givesi2}
\cos i= \frac{\delta}{\alpha} \left( \left(1 - \frac{B\delta}{y}\right)^{-4/3} - 1 \right)^{-1/2}.
\end{equation}
The mass of the BH can now be calculated by inserting this value of $i$ in 
(\ref{vzero}) for $\beta=0$. We find
\begin{equation}\label{givesm}
M_{\bullet} = \frac{\alpha d}{G} \left(\frac{y}{D \sin i} \right)^2 .
\end{equation}

We have checked the usefulness of these formulae by using them to analyse
the realistic spectrum shown in the bottom panel of Fig.~\ref{all}.  The
caustic occurs at the fourth pixel along the slit from the nucleus, and is
located at the seventh pixel in the dispersion direction. These numbers
yield $i=50\degr$ and $M_\bullet =2.6 \times 10^8\msun$ compared to the input
values $i=60\degr$ and $M_\bullet=10^8\msun$. The errors here are dominated
by the neglect in the above derivation of the stellar mass distribution which
is unknown. They can be reduced by recalling that the position of the caustic 
along the slit, $\alpha$, is fairly independent of the stellar density profile,
but its position in the dispersion direction, $y$, is sensitive to the
stellar density profile because it measures the mass interior to $\alpha$.
The calculations above should have used the position of the caustic
$y_\bullet$ when the BH acts alone.  From equation (\ref{givesy}) we
have that $y/y_\bullet=v_0(\alpha)/v_{\bullet}(\alpha)$, where
$v_{\bullet}(\alpha)$ is the velocity due to the BH alone.  Let 
$v_*(\alpha)$ be the velocity due to the stars, then
 \begin{equation}
y = y_{\bullet} \sqrt{1+\left(\frac{v_*}{v_{\bullet}}\right)^2}.
\end{equation}
Setting $v_*/v_{\bullet} = 1$ (see Figs.~\ref{rotcurve} and \ref{shifts}), 
we get $i=66\degr$ and $M_\bullet=0.86 \times 10^8\msun$ in good agreement 
with the input values.

The key advantage of this technique for determining $M_\bullet$ is that it
exploits a feature that occurs at the outer edge of the BH's sphere of
influence, and therefore gives higher sensitivity to BH detection than the
traditional method, in which one fits the Keplerian rise of the rotation
curve at the smallest accessible radii. Moreover, our approach yields the
mass of the black hole and the disc inclination independently, whereas the
traditional approach only yields $M_{\bullet} \sin^2i$.
One should keep in mind though that our method yields the mass within the 
radius of the caustic, and to get the BH mass one must assume the contribution 
to the potential from extended mass. This method cannot determine whether
the measured mass comes from the BH, but on the other hand the very existence
of the caustic comes from the steep velocity rise towards the nucleus,
which itself is characteristic of a massive BH. Thus the main advantage of
our technique is in detecting BHs in galaxies for which we cannot
achieve the resolution required to follow the Keplerian rise in velocities
inwards.

\section{Inclined slits}

Since nuclear discs frequently lie in a plane that differs
markedly from that of galaxy's main disc, one usually does not know the
line of nodes before nuclear spectra are taken. Hence, we now consider a
slit that passes through the nucleus at an arbitrary position angle (PA) 
with respect to the line of nodes.

\begin{figure}
\vspace{5mm}
\centerline{\psfig{file=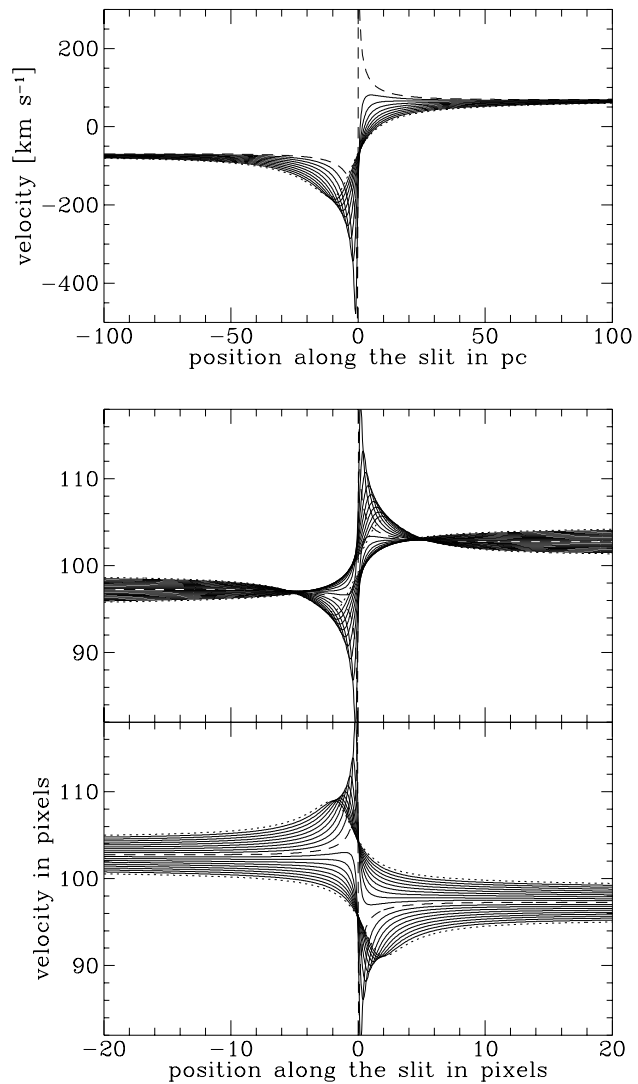,width=1.35\hsize}}
\vspace{-5mm}
\caption{The effect of non-zero angle ($\hbox{PA}=40\degr$) between the slit
and the line of nodes. Top panel: the position-velocity diagram for the
upper half of the slit, as in Fig.~\ref{larger} -- velocities for the lower
half can be obtained by inverting the diagram through its centre. Middle
panel: the distribution of light on the spectrograph's detector, as in
Fig.~\ref{all}. Bottom panel: the same but for a disc that rotates in the
opposite sense.\label{pa}}
\end{figure}

The top panel of Fig.~\ref{pa} is the analogue of Fig.~\ref{larger} for the
case of non-zero PA: it is a position-velocity plot for several cuts
that are inclined at $\hbox{PA}=40\degr$ to the line of nodes. The dashed
curve is for the cut that passes through the nucleus, and the other cuts all
pass above the nucleus. Notice that this plot, unlike Fig.~\ref{larger}, is
not symmetric on inversion through its centre; the plot one obtains by
carrying out this inversion is the position-velocity plot for cuts that pass
below the nucleus.

The middle panel of Fig.~\ref{pa} is the analogue of the top panel of
Fig.~\ref{all} for non-zero PA. It shows the effect of adding the
velocity offsets associated with finite slit width to the velocities shown
in the top panel, together with the corresponding velocities for the cuts
below the nucleus.  

The bottom panel is the analogue of the middle panel for a disc that rotates
in the opposite sense: thus before adding the velocity offsets we changed the 
sign of the velocities given by  the top panel.
Clearly the bottom and middle panels differ by much more than a
reflection. In particular, one has prominent caustics on the spatial axis
while the other does not.  Given that these two panels describe observations
with the same spectrograph of systems that differ only in their sense of
rotation, the striking differences between these panels may be surprising.
They reflect the fact that a spectrograph is symmetric only under inversion
of both the spatial and temporal axes, and not symmetric under time reversal
only. The latter asymmetry is evident because a spectrograph displays
frequency shifts as spatial displacements. For the same reason, changing 
the sense of rotation of the disc is equivalent to switching the sign
of the observed spectral order (see eq.~\ref{givest0}).

When the slit lies along the line of nodes, cuts that pass $\pm\delta$ 
above and below the slit map to the same
point in the position-velocity plot at a given distance down the
slit (see Fig.~\ref{larger}). Consequently,
it is immaterial whether contributions coming from above the nucleus are
moved up and the others down, or vice-versa. When the slit does not run
along the line of nodes, cuts above and below the nucleus map to different
points in the position-velocity plot, and it matters which is moved up and 
which down. 

Although the appearance of the spectrum changes with the slit position
angle, the principal features persist. In the middle and bottom panels of
Fig.~\ref{pa}, one can see the caustics, and two intensity maxima inside
the caustics. Notice that, whereas in Fig.~\ref{all} the outer envelope of
the light distribution near the nucleus
is formed by light coming through the slit
centre, in Fig.~\ref{pa} it comes from the sides of the slit -- the dashed
line associated with the slit centre has a much narrower peak.
Consequently, fitting profiles from an infinitely thin slit to the light
distributions in the lower panels
of Fig.~\ref{pa} would result in a considerable overestimation
of the BH mass when the traditional method is being used. We give an example 
of this phenomenon in Section 6.

Our method relies on the existence of caustics to diagnose the 
presence of a BH. Since Fig.~\ref{pa} shows that for non-zero PA the
appearance of caustics can depend on the sense of disc rotation, it follows
that the detectability of a BH with a given spectrograph setup depends on the 
sense of rotation of its accretion disc. As the slit position angle increases, 
the caustic becomes more difficult to detect as it moves in to the nucleus and  
often to higher velocities.

With unknown PA, our method returns only two out of three parameters: PA,
$i$, $M_\bullet$. However, equations (\ref{givesi}) to (\ref{givesm}) are 
readily generalized to the case of non-zero position angle.
Equation (\ref{vzero}) gives the line-of-sight
velocity at a point $(\alpha,\beta)$ in the disc, with $\beta=0$ marking
the line of nodes through the nucleus. For a cut inclined at
a given PA, the coordinate $\alpha'$ down the slit, and the offset $\beta'$
of the cut are related to $(\alpha,\beta)$ by a simple rotation matrix
\begin{eqnarray}\label{rotmat}
\alpha & = & \alpha' \cos \PA - \beta' \sin \PA \\
\beta  & = & \beta'  \sin \PA + \beta' \cos \PA .
\end{eqnarray}
Substituting these values of
$(\alpha,\beta)$ into equation (\ref{vzero}) gives the
velocity $v_0$ in the nuclear cut ($\beta'=0$), and $v_{\pm \delta}$ in the 
cut marking the edge of the slit ($\beta'=\pm \delta$).
As in (\ref{givesi}), the inclination angle can be recovered from
the ratio $v_0/v_{\pm \delta}$:
\begin{equation}\label{ifrompa}
\cos^2i = \frac{\Gamma^{4/3} (\sin \PA \pm \frac{\delta}{\alpha} \cos \PA)^2  
- \sin^2 \PA}
  {\cos^2 \PA - \Gamma^{4/3} (\cos \PA \mp \frac{\delta}{\alpha} \sin \PA)^2 },
\end{equation}
 where
 \begin{equation}\label{gamma}
 \Gamma = \frac{v_{\pm \delta}}{v_0} \
\left(1 \mp \frac{\delta}{\alpha} \tan \PA\right)^{-1} , 
\end{equation}
 and the prime has been dropped from $\alpha$ for simplicity. In equations
(\ref{ifrompa}) and (\ref{gamma}), the upper sign refers to the top edge of
the slit, and the lower sign to the bottom edge -- this dichotomy reflects
the loss of the symmetry that takes place as the PA is displaced from zero.
From equation (\ref{givesy}) we again get $v_{\pm \delta}/v_0 = 1 \mp
\frac{B\delta}{y}$. Substituting this ratio into equation (\ref{gamma}) we
obtain an expression for the inclination in terms of observed quantities.
The mass of the black hole can then be obtained from equations (\ref{vzero})
and (\ref{rotmat}) with $\beta'=0$:
 \begin{equation}\label{mfrompa} 
M_\bullet \; = \; M_\bullet(\PA=0) \; 
\frac{\left(1- \cos^2 \PA \, \sin^2 i\right)^{3/2}}{\cos^3 i \, \cos^2 \PA} ,
\end{equation}
where $M_\bullet(\PA=0)$ is given by equation (\ref{givesm}). Note that the
correction factor in the equation above quickly diverges for $i \rightarrow
90\degr$ and PA $\rightarrow 90\degr$, but otherwise it remains a mild
function of $i$ and PA.
In Section 6, we use published data and these formulae   to obtain a
new estimate of  the mass of the BH in M84.

\subsection{Offset Slits}

\begin{figure}
\centerline{\psfig{file=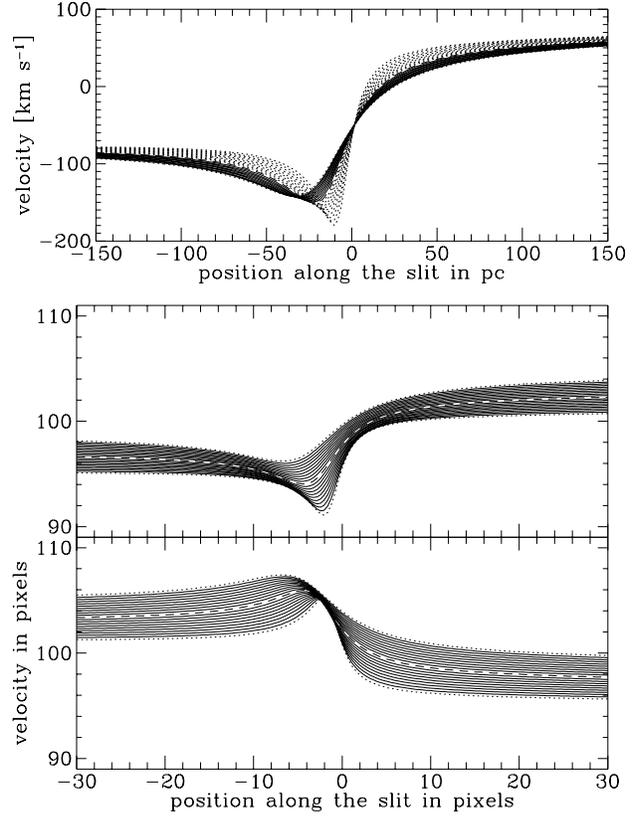,width=\hsize}}
 \caption{The same as Fig.~\ref{pa} but with the centre of the slit
displaced by the slit width, $0.2\,$arcsec, from the nucleus. The outer dotted
lines sample the slit edges, the dashed ones the slit centre. The top panel
is a position-velocity plot, with additional set of dotted lines sampling
the bottom half of the slit. The middle
and bottom panels show the light distributions in the spectrograph's
detector for discs rotating in opposite senses.\label{offcenter}}
\end{figure}

It is a common strategy to use offset slits to recover the PA of the 
nuclear disc, and to disentangle values of
$M_\bullet$ and $i$ from the value of the product $M_{\bullet} \sin^2 i$
that emerges from the conventional treatment of a slit through the nucleus.
We have shown how this disentanglement can already be done directly
from data for a wide slit through the nucleus, and it is a priori evident 
that a proper treatment of wide
slits greatly reduces the value of offset slits: a wide slit through the
centre already contains a narrow off-set slit. Constraints on the disc PA
can be inferred from offset slits, and there are a
couple of features of offset slits that merit mention.

First, the pattern of light on the detector is the same for slits that
pass equal distances above and below  the nucleus. Fig.~\ref{offcenter} 
is the analogue of Fig.~\ref{pa} for an offset slit -- 
the  PA is still $40\degr$ but now the centre of the slit misses the nucleus
by $0.2\,$arcsec. Again we see that the
light distribution in the detector differs  qualitatively depending on the
sense of rotation of the disc. The bottom  boundary of the light 
distribution in the two lower panels of Fig.~\ref{offcenter} is set by 
light from the edge of the slit that passes closest to the nucleus. The 
nuclear BH causes a mere widening of the light pattern close to the galactic 
centre in the middle panel, while in the lower panel a caustic is evident.
This panel corresponds to the same sense of rotation as the bottom panel in
Fig.~\ref{pa}, in which there is no comparable concentration of light.
Hence, when the disc rotates in the least favourable sense for black hole
detection with a slit through the nucleus, this is the favoured sense of
rotation for detection from an offset slit.

Finally, fitting the thin-slit model to the off-center data again results in an
overestimate of the BH mass: The disc's intensity will be highest along the
edge of the slit that passes closest to the nucleus.  In the two lower panels
of Fig.~\ref{offcenter} this edge forms the lower dotted line. Since the
curvature of this line is larger than the curvature of the dashed central
line, the BH mass will be overestimated if the curvature of the light
profile serves as the BH-mass estimator.

\section{Convolution with the PSF}

The discussion above has been confined to the distribution of the principal
maxima in the diffraction pattern that is produced on the spectrograph's
detector by light of a given frequency coming from a given point in the
object being measured. In reality light from a point on the target is
distributed over the slit by the PSF of the primary telescope optics, and
light from any point on the slit produces an entire diffraction pattern on
the detector.
To what extent would our conclusions be modified if
we took into account the complete diffraction pattern?

We cannot answer this question rigorously for lack of sufficient data on
STIS. To see what data are required, recall that light at a given frequency
from a given point on the object is concentrated by the primary telescope
optics into an interference pattern in the plane of the slit.  Ideally,
this pattern would comprise an Airy disc and rings, but it is in reality
complex.  The slit truncates this pattern. After passage through the
spectrograph, light from every surviving point in the pattern interferes
both with itself and light from other surviving points in the pattern, so 
in principle one needs to know the phase of the radiation within the slit 
and not just the intensity pattern. 

Here we neglect the effects of the coherence of radiation at different 
points in the slit, and add contributions from truncated Airy discs in 
intensity -- this is strictly correct only in the case of a seeing-dominated 
PSF. For each point in the source we approximate the 
Airy pattern in the slit plane with the circular model PSF. We chose the PSF
obtained by
Roeland van der Marel (2000, private communication) from tests on STIS
that involved shifting a star across a 0.1 arcsec-wide slit. This PSF
is approximated by four Gaussians
\begin{equation}\label{psf}
{\rm PSF}(r) = 
\sum_{i=1}^{4} \frac{\gamma_i}{2 \pi \sigma_i^2} \exp (-r^2/2 \sigma_i^2),
\end{equation}
with parameters listed in Table 1. It has a FWHM of 0.057 arcsec. Its 
pattern is truncated by the slit edges. The light from each surviving point 
on a given Airy disc is propagated to the detector according to equation 
(\ref{givest1}), \ie it is placed at the maximum of the diffraction pattern.
This approach recovers the flux on the detector and the velocity
offset measured in the abovementioned set of tests, 
and therefore gives a faithful approximation of the light pattern 
on the detector. In the next section, we confront our models,
already convolved with the PSF, with the STIS data.

\begin{table}
\caption{Parameters of PSF model for STIS}
\begin{tabular}{ccc} \\
\hline
$i$   & $\gamma_i$ & $\sigma_i$ (arcsec) \\
\hline
 1 & 0.219954 & 0.019070 \\
 2 & 0.599261 & 0.041687 \\
 3 & 0.147840 & 0.172908 \\
 4 & 0.032945 & 0.580370 \\
\hline
\end{tabular}
\end{table}

\section{Confronting observations}

\begin{figure*}
\centerline{\psfig{file=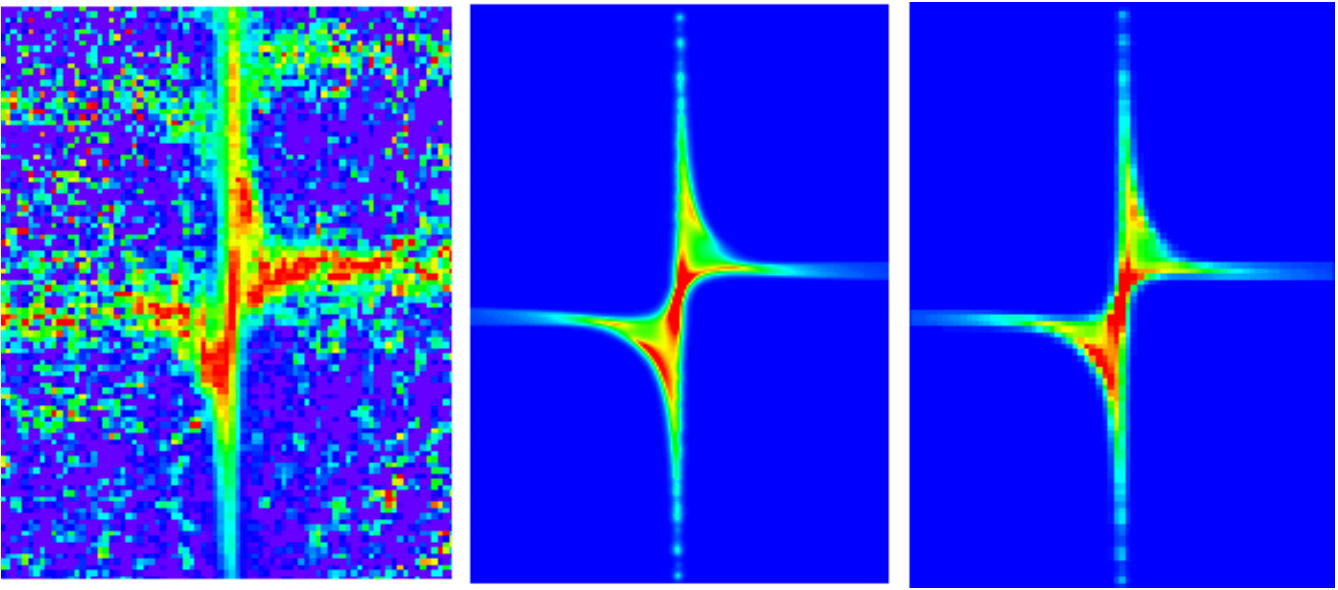,width=\hsize}}
\caption{Left: an $H\alpha$ emission-line spectrum of M84  taken with
 STIS using the G750M grating (from Bower et al.\ 1998). The dispersion
direction is vertical. Centre: a modeled light distribution on the 
spectrograph's detector (convolved with instrumental PSF) for $i=74\degr$, 
$\hbox{PA}=21\degr$, $M_\bullet=4 \times 10^8\msun$, and the centre of the 
slit shifted by 0.03 arcsec from the nucleus. The distance to M84 has been 
taken to be $ 17\Mpc$ and the surface-brightness of the emission has been 
assumed to be proportional to $R^{-1.5}$. Right: light distribution from the
central panel integrated over detector pixels.
\label{bower}}
\end{figure*}

To date, the most unambiguous gas kinematical 
detection of a nuclear BH with STIS is that for 
M84 by Bower et al.\ (1998). The left panel of Fig.~\ref{bower} shows 
the relevant spectrum, which was taken with a slit $0.2\,$arcsec wide. At
several positions just to the right of the nucleus, the emission clearly
peaks at two velocities. Consequently, when Bower et al.\ fitted Gaussians
in velocity to the data, they found two Gaussians to be required at many
spatial locations. Bower et al.\ interpreted this finding in terms of two
gas components. Fig.~\ref{all} shows that when the non-negligible width of
the slit is taken into account, two peaks arise naturally from the simplest
model of the accretion disc, in which there is a unique velocity at a given
radius.  

In the left panel of Fig.~\ref{bower} we interpret the point --
$\sim0.45\,$arcsec (nine pixels) 
from the nucleus -- at which the track of maximum light
splits into two, as the position of the caustic. We estimate the
corresponding velocity to be $125\kms$. Adopting the Bower et al.\ value of
the relative position angle of the slit with respect to the line of nodes 
($\hbox{PA}= 21\degr$), and roughly accounting for the
stellar mass contribution, we derive $i=74\degr$ and
$M_{\bullet} = 4 \times 10^8\msun$ from equations (\ref{ifrompa}) and
(\ref{mfrompa}).  Our estimate of $i$ agrees roughly with that of Bower et
al.\ ($i=80\degr$), but our value for $M_\bullet$ is almost 4 times smaller
than that of Bower et al.  We believe the Bower et al.\ value is too large
because it is based on fitting a thin slit model to the outer envelope in
light distribution, which is dominated by light from
the edge of the slit.  In Section 4 we showed how this procedure 
can lead to a significant overestimation of $M_\bullet$.

The observed spectrum is distinctly left-right asymmetric, which makes it
different from a generic model in Fig.~\ref{all}.  Any failure
to get the centre of the slit to pass through the nucleus will inevitably
introduce such asymmetry. We therefore simulated a STIS spectrum for
$M_\bullet=4 \times 10^8\msun$, $i=74\degr$ and relative $\hbox{PA} = 21\degr$,
and the centre of the slit displaced from the nucleus by $0.03\,$arcsec,
less than a quarter of its width. The middle panel of Fig.~\ref{bower}
shows the light distribution on the detector after convolution with the
PSF; in the right panel, this light has been binned into
pixels. The simulated spectrum clearly reproduces the 
main features of the observed spectrum next to it: on the left-hand 
side of this spectrum, the BH-related rise in velocities is clear, with not 
much emission in the second light maximum at low velocities. On the other 
hand, most of the light on the right of the nucleus is gathered into the 
low-velocity maximum, with the emission from the BH-related velocity rise 
enhanced only close to the galactic centre. All emission recorded in the 
nuclear slit
spectrum observed by Bower \etal can be explained as coming from a thin disc
in circular motion around the central BH. Note that although the distribution 
of intensities changes when the  slit is displaced, the relative position 
of caustics remains unchanged. 

It is also worth noting that the convolution with PSF does not introduce 
any significant change in the light pattern on the detector already outlined 
in Fig.~\ref{all}. It is so because
the slit width considered here (0.2 arcsec) is substantially larger than 
the FWHM of the instrumental PSF. Thus in observations with a sufficiently
wide slit, the convolution with the PSF can be treated as a small correction 
to the slit effects described in this paper.

Since our analysis of spectra taken with a wide slit is motivated by a
program of STIS observations, we have confronted our simulated spectra with
the STIS observations of Bower et al.\ (1998). However, the effects of a
wide slit already may be evident in older spectra of M87 that were taken with
the Faint Object Camera on HST (Macchetto \etal 1997). Two bright spots 
located symmetrically on each side of the nucleus and separated by a drop 
in the emission in the innermost $0.1\,$arcsec may indicate the presence 
of caustics, but we cannot make a strong case here because of limited
quality of the data.

\section{Conclusions}

We have analysed the consequences of using a long-slit that is significantly
wider than the PSF of the primary telescope optics. When the object under
study contains large velocity gradients, such that the object's velocity can
change appreciably across the slit, the pattern of light that will be
measured on the detector differs qualitatively from what one would observe
with an ideal narrow slit. There is more information in a spectrum
taken with a wide slit that in a spectrum of equal S/N that is taken with a
narrow slit. This happy state of affairs in natural intuitively, since a
wide slit contains many narrow slits within it.

Competition between real velocity differences
within the object, and offsets in velocity that arise because light from
different points across the slit enter the spectrograph at different angles,
frequently cause the intensity of light on the detector to have more than one
peak at a given spatial location, even for a trivial velocity field in the
object. Such multiple peaks in spectra of galactic nuclear discs have already
been observed from the ground (\eg Rubin, Kenney \& Young 1997), and are
evident in STIS spectra (Bower \etal 1998). In this last observation, they
were interpreted as evidence for multiple components along a single line 
of sight. Bertola \etal (1998) showed that this does not need to be the case,
and serious physical problems are likely to be encountered when a 
three-dimensional model is sought that displays such components. Here
we confirm this finding, and in addition conclude that, when correctly 
interpreted, the multiple peaks have considerable diagnostic power: they 
enable us to detect nuclear black holes by exploiting data for a given galaxy
at larger distances from the nucleus than in the traditional approach to BH
mass-estimation. Our BH-detection method is therefore reliable down to lower 
black-hole masses ($M_\bullet$) for which detection of Keplerian rise of
velocity inwards is beyond the resolution limit.
Moreover, our estimator recovers $M_\bullet$ and the inclination $i$ of the
disc independently from a single long-slit spectrum, whereas from the same
spectrum the traditional approach determines only the value of the product
$M_\bullet\sin^2i$ and requires an additional spectrum from an off-set slit
to extract $M_\bullet$ and $i$ separately from the product.
Nevertheless we ought to point out that our method serves only as a rough
indicator of a central mass concentration, and tells us nothing
about the mass distribution inside the caustic. The traditional method that
follows the Keplerian rise in velocities will secure the BH detection once
data with sufficient resolution become available.

When the slit is inclined at some angle to the line of nodes, the
distribution of light on the spectrograph's detector differs qualitatively
depending on the sense in which the disc rotates. Consequently, with a given
spectrograph setup, the detectability of a black hole depends on the sense in
which the disc surrounding it rotates. 

We applied our mass estimator to a published spectrum of M84 (Bower \etal
1998), and find that in this galaxy $M_\bullet$ is 4 times smaller than
previously estimated. We argue that interpretation of data taken with a wide
slit in terms of an infinitely thin slit will cause $M_\bullet$ to be
systematically overestimated.  Currently we are estimating the masses of
nuclear BHs in a sample of $\sim50$ nearby galaxies for which we have STIS
spectra (Marconi \etal 2000, Axon \etal 2000).

\section*{Acknowledgment}
We would like to thank Gary Bower for providing us with the data shown in
Fig.~\ref{bower}  and Roeland van der Marel for providing us with
unpublished test data for STIS, and for useful comments.

\end{document}